\documentstyle[preprint,aps]{revtex}

\begin{document}
\title{Far-infrared c-axis conductivity of flux-grown Y$_{1-x}$Pr$_{x}$Ba$_{2}$Cu$%
_{3}$O$_{7}$ single crystals studied by spectral ellipsometry}
\author{C. Bernhard, T. Holden, A. Golnik*, C.T. Lin, and M. Cardona}
\author{1) Max-Planck-Institut f\"{u}r Festk\"{o}rperforschung Heisenbergstrasse 1}
\date{5.4.2000}
\maketitle

\begin{abstract}
The far-infrared c-axis conductivity of flux-grown Y$_{1-x}$Pr$_{x}$Ba$_{2}$%
Cu$_{3}$O$_{7}$ single crystals with 0.2$\leq $x$\leq $0.5 has been studied
by sepectral ellipsometry. We find that the c-axis response exhibits
spectral features similar to deoxygenated underdoped YBa$_{2}$Cu$_{3}$O$%
_{7-\delta }$, i.e., a pseudogap develops in the normal state, the phonon
mode at 320 cm$^{-1}$ exhibits an anomalous T-dependence and an additional
absorption peak forms at T. This suggests that the T$_{c}$ suppression in
flux grown Pr-substituted crystals is caused by a decrease of the hole
content and/or by carrier localization rather than by pair breaking.

\end{abstract}

\newpage

Superconductivity above the liquification temperature of nitrogen was first
discovered in 1987 in the cuprate high T$_{c}$ compound YBa$_{2}$Cu$_{3}$O$%
_{7-\delta }$ (Y-123) with a critical temperature of T$_{c}$=92 K \cite{Wu1}%
. It was subsequently shown that chemical substitution of Y by most
rare-earth elements as well as by La leaves T$_{c}$ almost unaffected with T$%
_{c}$%
\mbox{$>$}%
90 K \cite{Maple1}. Ever since, the reason why Pr-123 does not become
superconducting (SC) has been a puzzle \cite{Soderholm1}. A number of models
have been proposed which explain the absence of SC in Pr-123 either in terms
of pair-breaking by the magnetic Pr-moments \cite{Guo1}, a hole depletion of
the CuO$_{2}$ planes caused by a mixed valency of Pr$^{3+/4+}$ \cite
{Dalichaouch1} or a partial substitution of Pr$^{3+}$ for Ba$^{2+}$ \cite
{Taekawa1}, a modification of the charge transfer between CuO chains and CuO$%
_{2}$ planes \cite{Khomskii1}, or a hole redistribution and localization due
to a strong hybridization of the Pr 4f and O 2p orbitals \cite{Fehrenbacher1}%
. Recently, it has been reported that Pr-123 can also be made SC with T$_{c}$%
\mbox{$>$}%
90 K \cite{Zou1}. This finding has led to renewed interest in the electronic
properties of the Pr-123 compound. Yet the essential mechanism which decides
whether Pr-123 is a superconductor with T$_{c}$%
\mbox{$>$}%
90 K or a non SC insulator is still unknown.

In this paper we report ellipsometric measurements of the far-infrared
(FIR)\ c-axis conductivity of partially Pr-substituted flux grown Y$_{1-x}$Pr%
$_{x}$Ba$_{2}$Cu$_{3}$O$_{7-\delta }$ crystals with 0.2$\leq x\leq $0.5. We
show that the c-axis conductivity of the Pr-substituted crystals exhibits
virtually the same spectral features as deoxygenated and thus underdoped YBa$%
_{2}$Cu$_{3}$O$_{7-\delta }$ crystals, i.e. a spectral pseudogap forms in
the normal state, the oxygen bond-bending phonon mode at 320 cm$^{-1}$
exhibits an anomalous T-dependence, and an additional broad absorption peak
forms at low T. This finding implies that a similar effect causes the T$_{c}$
suppression in flux grown Pr substituted and deoxygenated Y-123 crystals,
namely a depletion and/or a localization of the Zhang-Rice-type hole
carriers of the CuO$_{2}$ planes. From our measurements we do not obtain any
decisive information about the hole depletion mechanism in Pr-substituted
samples.

The growth of the Pr-substituted Y$_{1-x}$Pr$_{x}$Ba$_{2}$Cu$_{3}$O$_{7}$
crystals has been described previously \cite{Lin1}. Farily large crystals
with a typical size of the ac-face of 3 by 0.5-1 mm have been used. They
have been annealed in an O$_{2}$ gas stream for 1 day (d) at 600 $%
{{}^\circ}%
$C then cooled within 1d to 400 $%
{{}^\circ}%
$C and further annealed for 10 d at 400 $%
{{}^\circ}%
$C. A Zn-substituted YBa$_{2}$Cu$_{2.94}$Zn$_{0.06}$O$_{6.95}$ crystal has
been annealed under similar conditions but with the final step at 500 $%
{{}^\circ}%
$C for 5d. The midpoint SC transition temperature T$_{c}$ and the 10 to 90\%
halfwidth $\Delta $T$_{c}$ have been determined by dc-SQUID magnetisation
measurements at 5 Oe. The Pr-content and the Zn-content have been determined
by EDX analysis. For the Pr-substituted crystals we obtained T$_{c}$=81(3) K
for x$\approx $0.2, T$_{c}$=64(3) K for x$\approx $0.3, T$_{c}$=48(4) K for x%
$\approx $0.4, and T$_{c}$=24(5) K for x$\approx $0.5. For the
Zn-substituted crystal we obtained T$_{c}$=76(3) K for z$\approx $0.06.

The ellipsometric measurements have been performed at the National
Synchrotron Light Source (NSLS) using a home built ellipsometer attached to
a Nicolet Fast-Fourier spectrometer at the U4IR beamline \cite{Henn1}. Some
experiments have been done with the ellipsometer attached to a Bruker 113V
using a conventional Hg arc light source. The optical measurements have been
performed on the as grown clean \ ac surfaces (with the c-axis in the plane
of incidence). The technique of ellipsometry provides significant advantages
over conventional reflection methods in that (a) it is self-normalizing and
does not require reference measurements and (b) the real and the imaginary
parts of the dielectric function, $\varepsilon $=$\varepsilon _{1}$+i$%
\varepsilon _{2}$, are obtained \ directly without a Kramers-Kronig
transformation. Since only relative intensities of the reflected light are
required, the ellipsometric measurements are more accurate and reproducible
than conventional reflection measurements.

Figure 1 shows the spectra of the FIR c-axis conductivity, $\sigma _{1c}$,
of the Y$_{0.8}$Pr$_{0.2}$Ba$_{2}$Cu$_{3}$O$_{7}$ crystal with T$_{c}$
=81(3) K in the normal and in the superconducting state. The room
temperature spectrum is composed of an almost frequency independent
electronic background on which five infrared active phonon modes are
superimposed. The phonon modes correspond to those of fully oxygenated YBa$%
_{2}$Cu$_{3}$O$_{7}$ \cite{Henn2}. It has been previously shown for
deoxygenated YBa$_{2}$Cu$_{3}$O$_{7-\delta }$ that the strength of the
phonon mode at 630 cm$^{-1},$ which corresponds to the vibration of apical
oxygen neighboring an empty chain fragment, is a good indicator for the
oxygen deficiency of a given sample. The absence of the 630 cm$^{-1}$ phonon
mode thus confirms the fully oxidized state of our present crystal. With
decreasing T the electronic background can be seen to develop a spectral gap
in the normal state already well above T$_{c}$. The characteristic features
of this normal state spectral gap are identical to those of the pseudogap
which has been observed in an underdoped deoxygenated YBa$_{2}$Cu$_{3}$O$%
_{6.75}$ crystal with similar T$_{c}$ \cite{Bernhard1,Bernhard2}. In both
crystals the normal state pseudogap has a similar size of $\omega _{NG}\sim $%
700-800 cm$^{-1}$ (defined as the onset of the suppression of the
conductivity with decreasing temperature) and its onset temperature T$_{NG}$
is around 200 K. Even the spectral shape of the pseudogaps compares very
well. Close to T$_{c}$, the in-plane oxygen bond-bending mode at 320 cm$%
^{-1} $ becomes strongly renormalized, its spectral weight decreases and its
position is shifted by about 10 cm$^{-1}$ towards lower energies.
Simultaneously, an additional broad peak appears around 500 cm$^{-1}$ as
indicated by the arrow. Once more, the same spectral features have been
observed in the FIR c-axis response of underdoped deoxygenated Y-123 \cite
{Bernhard1,Bernhard2}. The broad low-T peak and the related strong anomaly
of the 320 cm$^{-1}$ phonon mode have been successfully explained in terms
of a model where the bilayer cuprate compounds like Y-123 are treated as a
superlattice of intra- and interbilayer Josephson junctions \cite
{Munzar1,Grueninger1}. Within this model, the broad low-T peak corresponds
to the transverse optical Josephson plasmon which arises from the out of
phase oscillation of the intra-and the interbilayer longitudinal plasmon
modes. The strong anomaly of the 320 cm$^{-1}$ phonon mode is explained as
due to the drastic changes of the local electric fields acting on the
in-plane oxygens as the Josephson currents set in the SC state \cite{Munzar1}%
. Overall, this close analogy implies that the T$_{c}$ suppression upon
Pr-substitution is caused by a depletion and/or a localization of the mobile
holes of the CuO$_{2}$ planes rather than by pair-breaking.

For comparison, we also investigated a YBa$_{2}$Cu$_{2.94}$Zn$_{0.04}$O$%
_{6.9}$ crystal whose T$_{c}$ is suppressed by a similar amount to T$_{c}$%
=76(3) K. Meanwhile, it is well established that the T$_{c}$-suppression in
Zn-substituted samples is caused by strong pair-breaking due to impurity
scattering in the unitarity limit on the Zn impurities while the hole
content of the CuO$_{2}$ plane is hardly affected \cite{Tallon1}. Figure 2
displays $\sigma _{1c}$ at different temperatures between room temperature
and 10 K. It is evident that no sign of a pseudogap in the normal state
c-axis conductivity occurs, despite that fact that the T$_{c}$ value is
severely suppressed. The c-axis conductivity instead hardly changes in the
normal state. If anything, it rather exhibits a very weak Drude-like
behavior since $\sigma _{1c}^{el}$ increases slightly with decreasing T and
towards low frequency. Note that a similar behavior has been observed in a
Zn-free YBa$_{2}$Cu$_{3}$O$_{6.9}$ crystal which was annealed under the same
conditions being almost optimally doped with T$_{c}$=92 K \cite
{Bernhard1,Bernhard3}. Also the anomaly of the 320 cm$^{-1}$ phonon mode is
much weaker for the Zn-substituted crystal than for the 20 \% Pr-substituted
crystal and there is no clear evidence for the additional low-T peak. For
the Zn-substituted crystal a spectral gap forms only in the superconducting
state below T$_{c}$=76 K. Then the pair breaking effect due to the
Zn-impurities is clearly evident. Firstly, the size of the spectral gap is
significantly reduced to 2$\Delta _{SC}\approx $450 cm$^{-1}$as compared to 2%
$\Delta _{SC}\approx $650 cm$^{-1}$ in pure optimally doped YBa$_{2}$Cu$_{3}$%
O$_{6.9}$ \cite{Bernhard1}. Secondly, the residual conductivity remains very
large even at the lowest temperature of 10 K. This indicates that a large
number of quasi-particles remain unpaired in the SC state due to the strong
pair-breaking effect of the Zn-impurities. No such characterstic signature
of the pair-breaking effect is observed for the Pr-substituted crystals.

Figure 3 shows the T-dependence of the FIR c-axis conductivity of the Pr$%
_{0.3}$Y$_{0.7}$Ba$_{2}$Cu$_{3}$O$_{7}$ crystal with T$_{c}$=64(3) K which
is more strongly Pr-substituted than the previous one. The normal state
spectra are shown in Fig. 3(a). Once more a spectral pseudogap can be seen
to develop in the normal state. The pseudogap now already starts to form
around room temperature and its size clearly exceeds the measured spectral
range, i.e. $\omega _{NG}$%
\mbox{$>$}%
700 cm$^{-1}$. A corresponding increase of the onset temperature and of the
size of the pseudogap with increasing underdoping of the CuO$_{2}$ planes
has been previously observed for strongly deoxygenated Y-123 crystals \cite
{Bernhard1,Bernhard2,Bernhard3}. All the characteristic spectral features
are very similar like in a deoxygenated Y-123 crystal with comparable T$%
_{c}\sim $60 K. On the other hand, the absolute value of the $\sigma
_{el}^{1c}$ is significantly higher in the Pr-substituted crystal. This
effect can be understood to be due to the presence of the fully oxygenated
and thus metallic CuO chains. Evidence for metallic CuO chains has been
obtained even in pure Pr-123 \cite{Takaneka1}. A similar chain-related
effect on $\sigma _{el}^{1c}$ has previously been observed for
Ca-substituted Y$_{1-x}$Ca$_{x}$Ba$_{2}$Cu$_{3}$O$_{7-\delta }$ crystals for
which the absolute value depends on the oxygen content and thus the
metallicity of the CuO chains while the characteristic frequency- and
T-dependence is determined mainly by the hole doping of the CuO$_{2}$ planes 
\cite{Bernhard1,Bernhard3}. Another difference as compared to the
deoxygenated Y-123 crystals is the observation of two weak defect modes in
the spectral range of 400 to 450 cm$^{-1}$. These modes are present already
at room temperature but become sharper and thus more pronounced at low T.
These defect modes possibly originate from oxygen defects within the CuO
chain layer other than the usual O(1) oxygen defects in Y-123.
Alternatively, they might corresond to crystal field excitations of the
magnetic Pr$^{3+}$ ions. Figure 3(b) shows the low temperature data in the
SC state. It can be seen that the 320 cm$^{-1}$ phonon mode becomes strongly
renormalized and that an additional peak develops around 420 cm$^{-1}$ again
in close analogy to deoxygenated underdoped YBa$_{2}$Cu$_{3}$O$_{6.6}$ with T%
$_{c}\sim $60 K \cite{Homes1,Bernhard2}. The additional peak is not quite as
pronounced as for the Pr-free underdoped Y-123 crystals. In the context of
the Josephson-plasmon superlattice model this difference can be explained as
due to the larger quasiparticle conductivity of the Pr-substituted crystal
which leads to a stronger damping of the transversal Josephson plasmon mode 
\cite{Munzar1}. As was noted above, this difference can be attributed to the
metallic conductivity of the CuO chains and the subsequently higher normal
state c-axis conductivity of the Pr-substituted crystals. Also due to the
presence of the fully oxygenated metallic CuO chains, the interbilayer
Josephson plasmon frequency, as deduced from the zero crossing of $%
\varepsilon _{1}$ and/or from the low frequency slope of $\varepsilon _{1}$,
appears to be somewhat higher in the Pr-substituted crystals compared to
deoxygenated Y-123 with similar T$_{c}$. More details about the dependence
of the inter- and intrabilayer Josephson plasmons on the presence of the
metallic CuO chains, including fits with the model given in \cite{Munzar1},\
will be presented in a forthcoming publication.

Figures 4 and 5 show the FIR c-axis conductivity of Pr$_{0.4}$Y$_{0.6}$Ba$%
_{2}$Cu$_{3}$O$_{7}$ with T$_{c}$=48(4) K and Pr$_{0.5}$Y$_{0.5}$Ba$_{2}$Cu$%
_{3}$O$_{7}$ with T$_{c}$=24(5) K, respectively. The absolute values $\sigma
_{1c}^{el}$ once more are significantly higher for the fully oxygenated
Pr-substituted crystals than for comparably underdoped deoxygenated Y-123.
This circumstance allows us to see more clearly the characteristic features
of the normal state pseudogap in such strongly underdoped 123-type samples.
For the x=0.4 crystal it is evident that the size of the normal state
pseudogap exceeds the measured spectral range by far, i.e., $\omega _{NG}$%
\mbox{$>$}%
\mbox{$>$}%
700 cm$^{-1}$. This finding confirms our previous report that the size of
the pseudogap of deoxygenated Y-123 crystals increases continuously on the
underdoped side \cite{Bernhard1,Bernhard2}. The pseudogap forms around room
temperature and it is not related to the formation of the additional low T
peak nor to the anomaly of the 320 cm$^{-1}$ phonon mode. Notably, the
characteristic shape of the anomaly of the 320 cm$^{-1}$ phonon mode and the
additional peak, which both merge for the present sample, indicate that the
size of the intrabilayer Josephson plasmon is somewhat smaller for this 40\%
Pr-substituted crystal than for a comparably underdoped pure YBa$_{2}$Cu$%
_{3} $O$_{6.5}$ crystal with T$_{c}$=52 K. Instead the spectral shape of the
anomalous 320 phonon mode and the additional peak resemble that of a more
strongly underdoped YBa$_{2}$Cu$_{3}$O$_{6.45}$ crystal with T$_{c}$=25 K 
\cite{Munzar1}. It worth noting that this effect is expected within the
Josephson superlattice model since the separation of the CuO$_{2}$ planes is
larger in the Pr-substituted crystal than in the pure Y-123 resulting in a
smaller intrabilayer Josephson plasma frequency for the Pr-substituted
sample. In contrast, as noted above, the interbilayer plasma frequency
appears to be larger in the Pr-substituted crystals due to the presence of
the metallic CuO chains. Figure 5 shows the FIR c-axis conductivity of our
most heavily Pr-substituted Pr$_{0.45}$Y$_{0.55}$Ba$_{2}$Cu$_{3}$O$_{7}$
crystal with T$_{c}$=24(5) K. It is remarkable that for this crystal, which
is located close to the metal insulator transition around x=0.55, the normal
state pseudogap suddenly is less pronounced and its size has decreased. This
finding seems to indicate that the pseudogap disappears around the metal
insulator transition. Note that it has not been possible to follow the
evolution of the electronic c-axis conductivity in such detail in
correspondingly underdoped deoxygenated Y-123 crystals for which the
absolute values of $\sigma _{1c}^{el}$ are much lower \cite{Munzar1,Homes1}.

Finally, we comment on the question whether our FIR c-axis conductivity data
provide any information about the mechanism which causes the hole depletion
of the CuO$_{2}$ planes upon Pr-substitution. Recent x-ray diffraction
measurements revealed that flux-grown Pr-123 crystals are Cu deficient on
the Cu(1) site and/or exhibit a partial substitution of Pr on the Ba-site 
\cite{Ye1}. It was argued that these effects may be responsible for the hole
depletion. The only anomalous features in the FIR phonon modes of our
Y,Pr-123 crystals which may be indicative of structural disorder within the
CuO chain layer are the weak defect mode at 420 cm$^{-1},$ whose origin is
yet unknown, and a broadening of the phonon mode at 280 cm$^{-1}$ due to
vibration of chain oxygen \cite{Henn1} which is more pronounced for the
Pr-substituted crystals than for deoxygenated crystals with similar T$_{c}$ 
\cite{Bernhard1,Bernhard2}. However, a similar defect mode around 400-450 cm$%
^{-1}$ and an even stronger broadening of the Cu(1) mode at 280 cm$^{-1}$
have recently been observed in Nd-123 which is superconducting with T$_{c}$%
=93 K \cite{Hauff1}. Therefore, while these features may be related to a
deficiency on the Cu(1) site, it is unlikely that they are responsible for
the strong T$_{c}$-suppression which occurs only for the Pr-substituted
crystal but not for Nd substituted one. In our spectra we do not observe any
anomalous broadening or shift for example of the phonon mode at 155 cm$^{-1}$
to which mainly Ba and Cu(1) contribute \cite{Henn1}. While the masses of Ba
and Pr are very similar one still expects this mode to be broadened if Ba$%
^{2+}$ is substituted by Pr$^{3+}$ since their electronic environment
(oxygen configuration) should be rather different \cite{Ye1}. To conclude,
we do not observe any anomalous changes of the FIR active phonon modes which
could be used as an indication that the decrease of the hole doping of the
CuO$_{2}$ planes can be explained as a chemical doping effect. Our data
rather favor the model of Fehrenbacher and Rice where the charge carriers of
the CuO$_{2}$ planes are redistributed and localized in hybridized Pr 4f and
O 2p orbitals \cite{Fehrenbacher1}. Note that this model is supported by
recent x-ray absorption measurements \cite{Merz1}.

In summary, by spectral ellipsometry we have studied the far-infrared c-axis
conductivity of Pr-substituted Pr$_{x}$Y$_{1-x}$Ba$_{2}$Cu$_{2}$O$_{7}$
crystals with 0.2$\leq $x$\leq $0.5. We have shown that the c-axis
conductivity of these crystals exhibits spectral features similar to
deoxygenated and thus underdoped YBa$_{2}$Cu$_{3}$O$_{7-\delta }$. A
spectral pseudogap already forms in the normal state, the oxygen
bond-bending phonon mode at 320 cm$^{-1}$ exhibits an anomalous T dependence
and an additional broad absorption peak forms at low T. This finding
suggests that the T$_{c}$ suppression in the Pr-substituted samples is
caused by a decrease in the concentration or a localization of the mobile
hole carriers of the CuO$_{2}$ planes.

We gratefully acknowledge G.P. Williams and L. Carr for technical help at
the U4IR beamline at NSLS and E. Br\"{u}cher and R.K. Kremer for performing
the SQUID magnetisation measurements.

* Permanent address: Institute of Experimental Physics, Warsaw University,
Ho\.{z}a 69, 00-681 Warsaw, Poland

\bigskip {\bf Figure Captions}

Figure 1: Spectra of the real part of the FIR c-axis conductivity of Pr$%
_{0.2}$Y$_{0.8}$Ba$_{2}$Cu$_{3}$O$_{7}$ with T$_{c}$=81(3) K in the normal-
and the superconducting state. The position of the additional low
temperature peak is indicated by the solid arrow.

\bigskip

\bigskip Figure 2: Optical FIR c-axis conductivity of Zn-substituted YBa$%
_{2} $Cu$_{2.94}$Zn$_{0.06}$O$_{6.9}$ with T$_{c}$=76(3) K at different
temperatures in the normal- and the superconducting state. The arrow
indicates the onset of the spectral gap in the superconducting state.

\bigskip

Figure 3: Temperature dependence of the real part of the FIR c-axis
conductivity of Pr$_{0.3}$Y$_{0.7}$Ba$_{2}$Cu$_{3}$O$_{7}$ with\ T$_{c}$%
=64(3) K, (a) in the normal state and (b) in the superconducting state. The
solid arrow marks the position of the additional low-temperature absorption
peak.

\bigskip

Figure 4: Real part of the FIR c-axis conductivity of Pr$_{0.4}$Y$_{0.6}$Ba$%
_{2}$Cu$_{3}$O$_{7}$ with T$_{c}$=48(4) K. The arrow indicates the position
of the low temperature absorption peak which merges with the phonon mode at
320 cm$^{-1}$.

\bigskip

Figure 5: Temperature dependence of the real part of the FIR c-axis
conductivity of Pr$_{0.5}$Y$_{0.5}$Ba$_{2}$Cu$_{3}$O$_{7}$ with T$_{c}$%
=24(5) K.

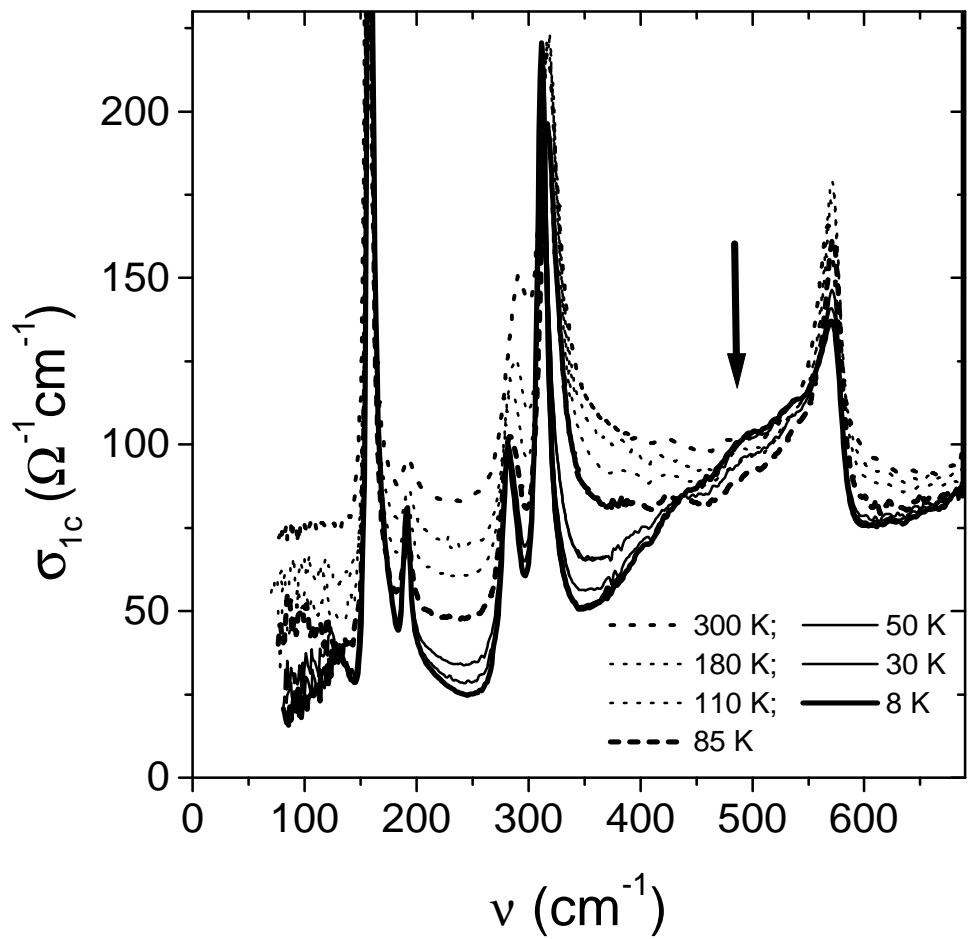
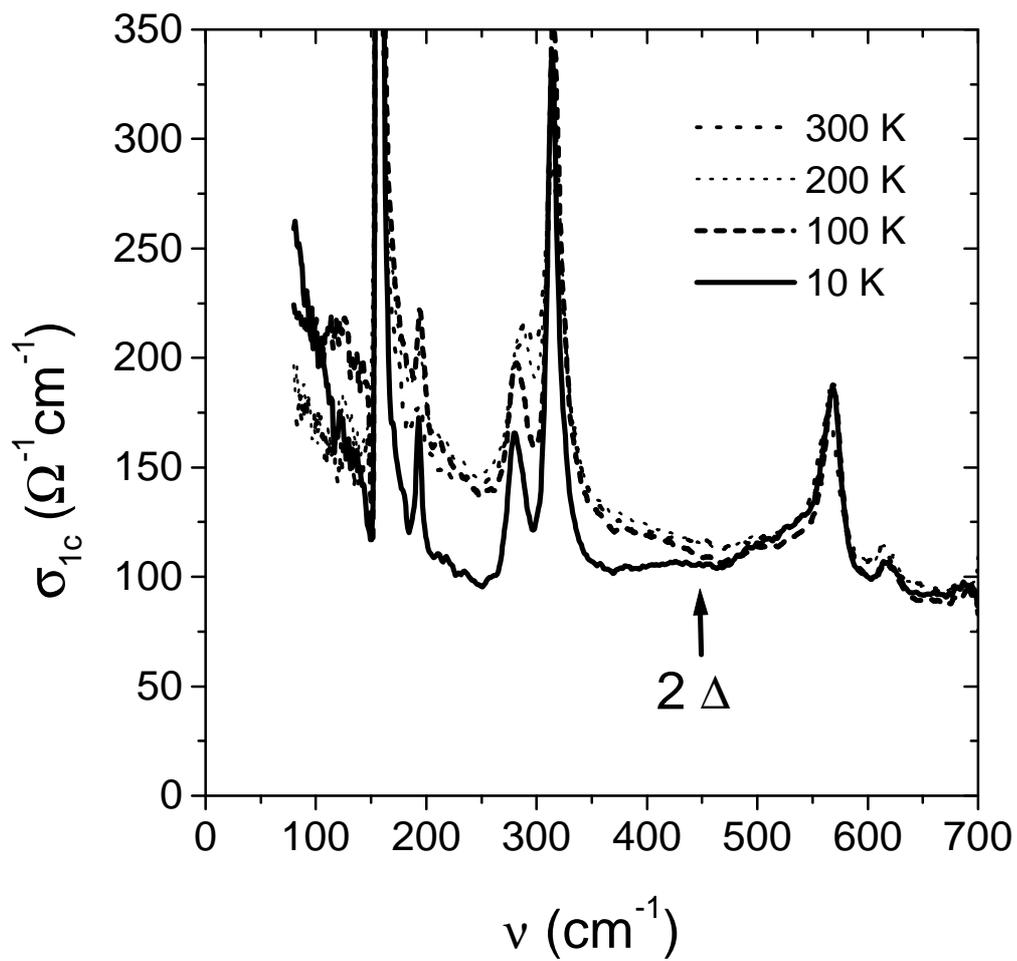
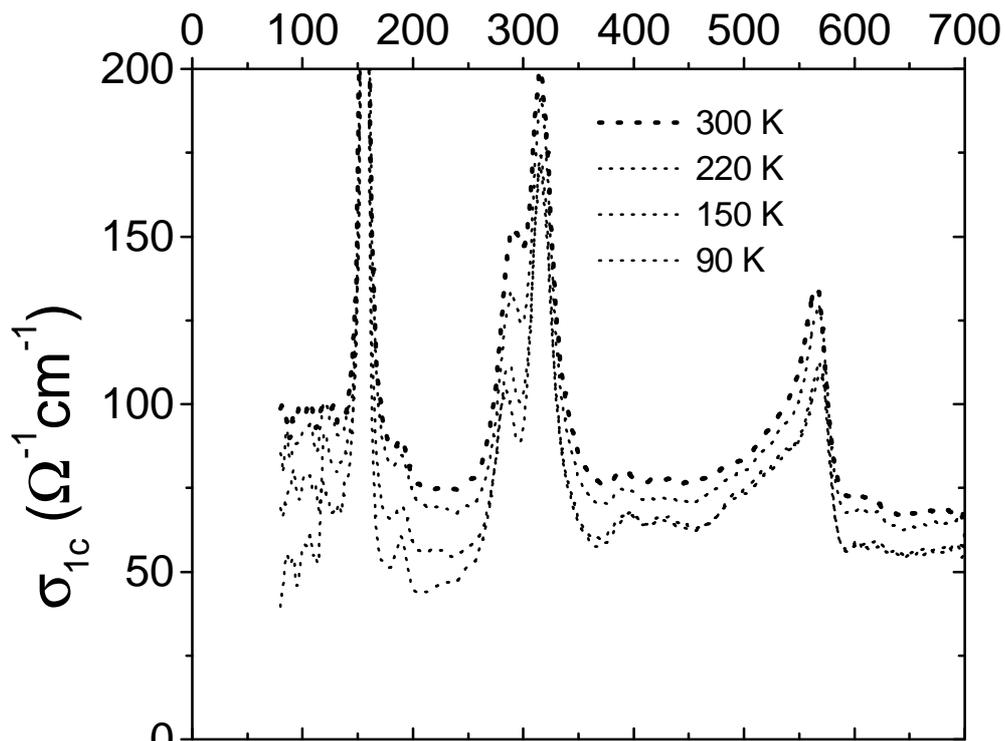
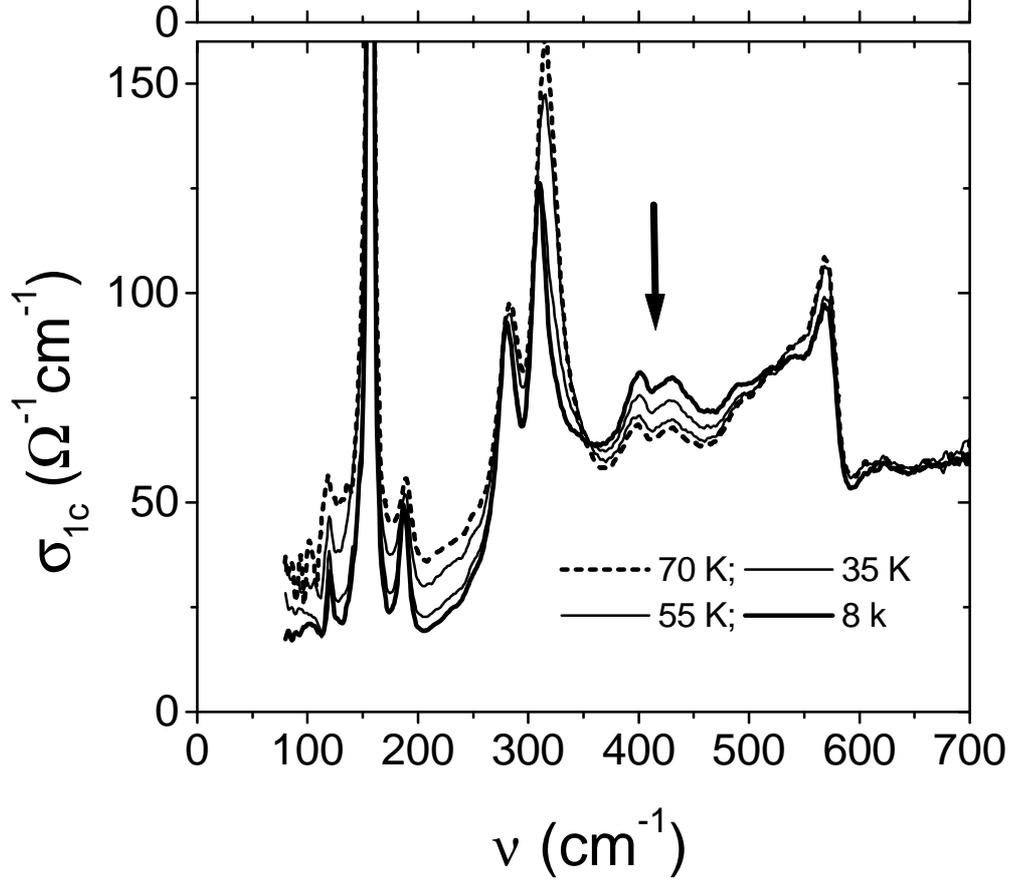
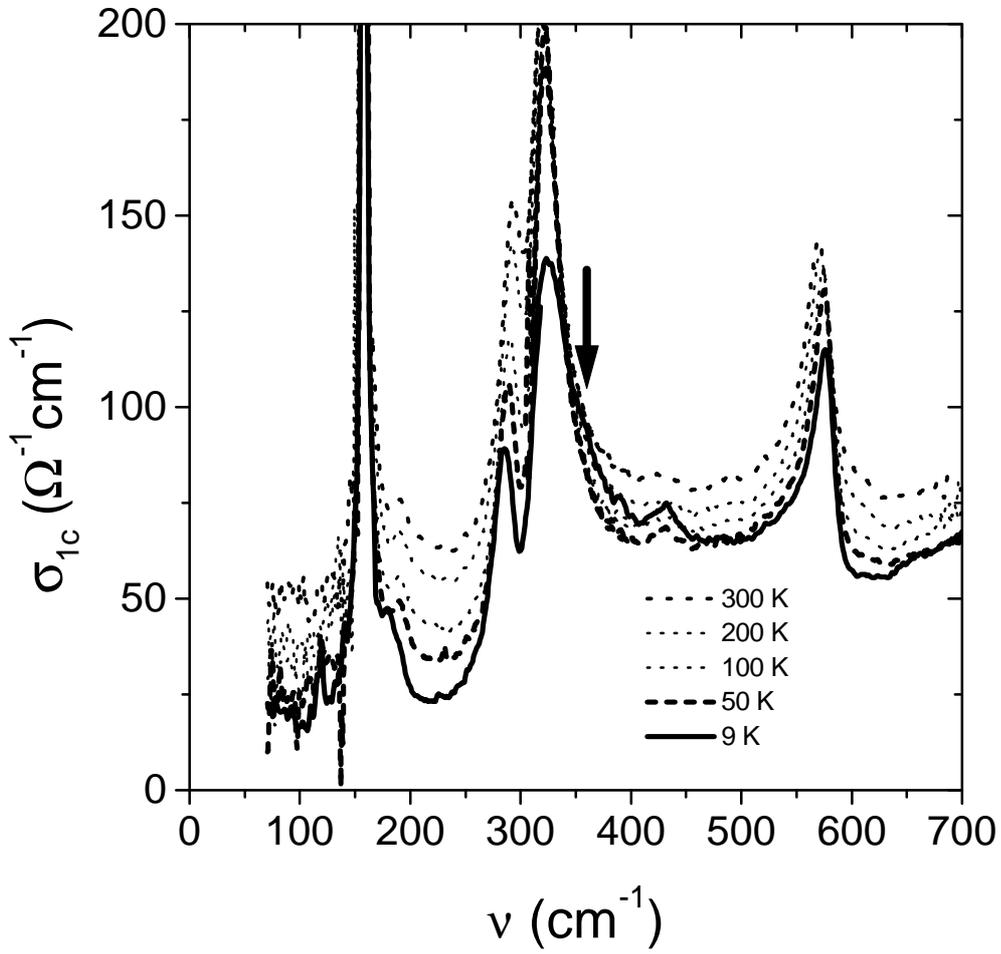
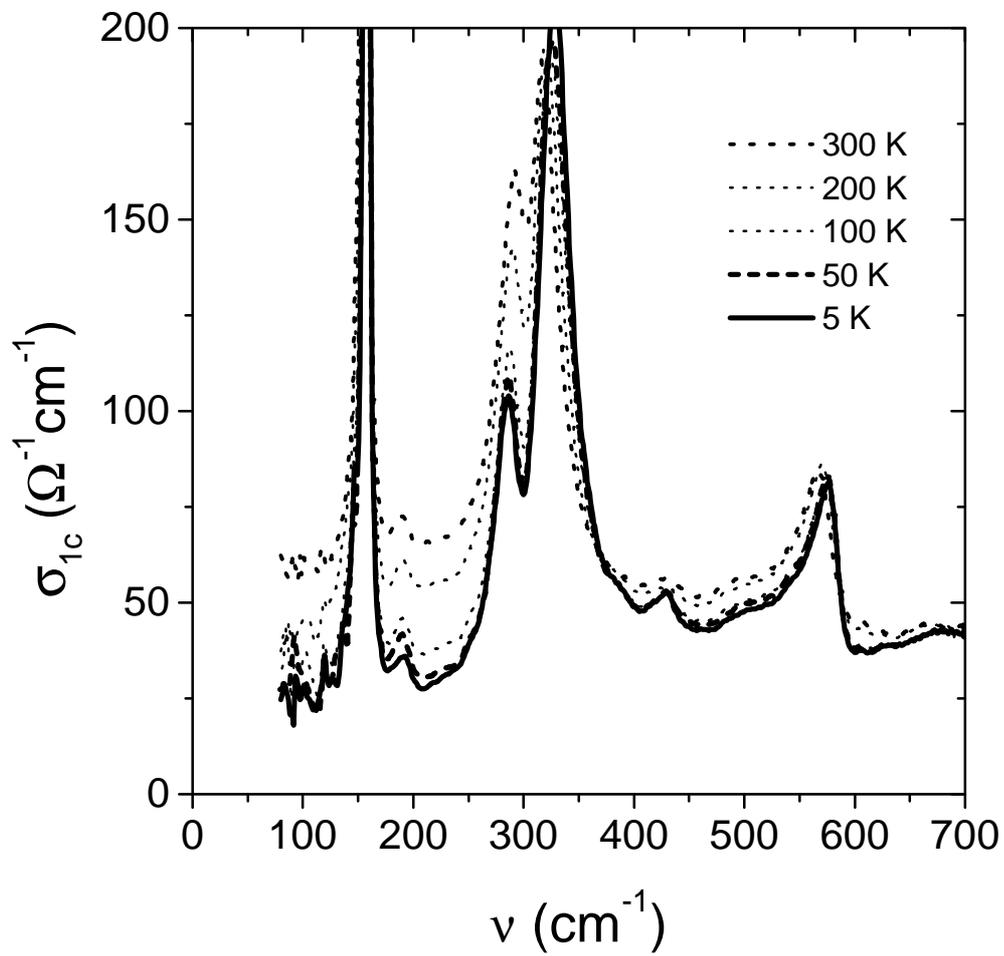
\end{document}